\documentstyle[11pt]{article}
\newcommand{\msun}{\, {\rm M}_\odot}

\newcommand{\nat}[2]{#1 \cdot 10^{#2}}
\newcommand{\rmin}{\, {\rm d}_{\rm min}}
\newcommand{\rmax}{\, {\rm R}_{\rm max}}
\input{psfig}
\psfigurepath{./}
\begin{document}
%
%
\begin{center}
{\Large \bf Modeling Encounters of Galaxies:}\\

 {\Large \bf The Case of NGC4449} \\
 
 {\sc Christian Theis}
 
{\it Institut f\"ur Theoretische Physik und Astrophysik, Universit\"at Kiel,
              24098 Kiel, Germany,
              email: theis@astrophysik.uni-kiel.de}
\end{center}
 
\unitlength1.0cm
\vspace*{-0.5cm}

 
%
\begin{abstract}

 Several N-body methods were combined in order to develop
a method for the determination of the parameters of interacting
galaxies. This method has been applied to the HI distribution
of NGC4449. 
In a first step the fast restricted N-body models were used
to confine a region in parameter space reproducing the main
observational features. In a second step a genetic algorithm
is introduced which allows for an automated search in the 
high-dimensional parameter space as well as for a uniqueness 
test of special parameter sets. Using artificial data I
show that the genetic algorithm reliably recovers
orbital parameters, provided that the data are sufficiently
accurate, i.e.\ all the key features are included.
  In the third step the results of the restricted N-body models
are compared with detailed self--consistent N-body simulations.
In the case of NGC4449 the applicability of the simple
restricted N-body calculations is demonstrated. Additionally,
it is shown that the HI gas can be modeled here without any restrictions
by a purely stellar dynamical approach.

   NGC4449 is an active star--forming dwarf galaxy of Magellanic type.
From radio observations van Woerden et al.\ (1975) found an extended 
HI--halo around NGC4449 which is at least a factor of 10 larger than the
optical diameter $D_{\rm 25} \approx 5.6$ kpc. More recently,
Bajaja et al. (1994) and Hunter et al.\ (1998) discerned details
in the HI--halo: a disc--like feature around the center of NGC4449 and
a lopsided arm structure. In a series of simulations it is demonstrated
that the main features can be obtained by a gravitational interaction
between NGC4449 and another nearby dwarf galaxy, DDO125. According to
these calculations the closest approach between both galaxies
happened $\sim \nat{3.5}{8}$ yr ago at a minimum distance of $\sim 25$ 
kpc on a nearly parabolic orbit. In case of an encounter scenario, 
the dynamical mass of DDO125 should not be smaller than 10\% of NGC4449's
mass. Before the encounter the observed HI 
was arranged in a disc with a radius of
35--40 kpc around the center of NGC4449 which had the same 
orientation as the central ellipsoidal HI structure.
The origin of this disc is still unclear, but it might be caused by a 
previous interaction.

\end{abstract}

%
%
\section{Introduction}

   During the last decades our picture of galaxies
being 'Welteninseln' (e.g.\ Kant, 1755), i.e.\ island universes, has
completely changed. At the beginning of extragalactic
research in the 1920s galaxies were thought to be
in general 'islands', i.e.\ isolated stellar systems, with just a
few obvious exceptions like the M51 system. The success of the Hubble
classification could have been a reason to discard systems not fitting
into it. Thus less regular
systems have been neglected  -- by definition -- as peculiar. The
interest in these systems raised strongly when new catalogues
by Vorontsov--Velyaminov (1959) and Arp (1966) became available.
They demonstrated that even the peculiar objects have a lot of
common features like bridges, tails, or rings.
Additionally, better observational instruments allowing for deeper images 
and new wavelength ranges revealed more complex structures. Thus,
a lot of galaxies formerly classified as non--interacting may have had
some gravitational interaction in the past which is still reflected
in e.g.\ warps, flares, or thick discs. Especially useful are 
HI observations which can cover a much larger radial extension than
the optical images. Therefore, HI is a very good tracer for tidal interactions.

  The theoretical understanding of interacting galaxies suffered
for a long time from the lack of computational power allowing for
a numerical solution of the gravitational N-body problem. After
a remarkable treatment by Holmberg (1941) who built an analogue computer
(consisting of light bulbs and photo cells) for determining the gravitational
force, it took 20 years until N-body simulations were performed on a general
purpose computer (Pfleiderer \& Siedentopf, 1961): The basic idea
of these {\it restricted N-body} simulations was the assumption that the
potential in interacting galaxies can adequately be modeled by
two particles which represent the galactic masses and move under their
mutual gravitation, i.e.\ on Keplerian orbits. With these
assumptions all the other particles are just test particles, and the 
complete N-body problem has been reduced to $N$ single body problems for a
time--dependent potential. In a remarkable series of simulations
Toomre \& Toomre (1972) applied this technique in order to determine
the parameters of a couple of well studied interacting
systems like Arp 295, M51 + NGC5195, or NGC4038/39. Motivated by
these simulations and by a remarkable coincidence of the estimated fraction
of strong interactions with the fraction of elliptical and S0 galaxies
Toomre (1977) put forward the idea of a merger origin of elliptical galaxies.
Hence, a whole class of galaxies would have been formed mainly by strong
interactions. However, restricted N-body simulations were not able to check
this idea, because they do not include processes like 
dynamical friction or transfer of orbital angular momentum of the galaxies into
galactic spin in a self--consistent manner. 
Moreover, the assumption of unaffected point--like
galaxies becomes rather poor for merging systems. Thus, computationally
much more expensive self--consistent calculations were required. Already
first models using $N=500$ particles proved the idea of merging 
(White, 1978; Gerhard, 1981). 

With increasing computational power
new N-body techniques have been developped which increased the accessible
particle number by many orders of magnitude. Especially the 
TREE--method developped by Barnes \& Hut (1986) and Hernquist (1987, 1990)
was perfectly suited for interacting systems, because the
organization of the force calculation -- the most time consuming part
of N-body calculations -- adapts in the TREE scheme
easily and automatically to clumpy mass distributions for a moderate
computational price ($\sim O(N\log(N))$) compared to direct simulations
($\sim O(N^2)$). By this, Barnes (1988) was able to simulate encounters
of disc galaxies including all dynamical components, i.e.\ the disc,
the bulge, and the halo as N-body systems.
Compared to faster grid--based methods (e.g.\ Sellwood, 1980)
or expansion methods (e.g.\ Hernquist \& Ostriker, 1992) direct N-body 
simulations (or semi--direct methods like TREEs) seem to be more flexible 
with respect to strongly varying geometries and scalelengths. 
An alternative to these sophisticated numerical techniques is the
use of special purpose computers like the machines of the GRAPE project 
(Sugimoto et al., 1990). They implement Newton's law of gravity
(modified for gravitational softening) in the hardware which allows for
a very fast direct determination of the gravitational forces, though
there still exists the $N^2$ bottleneck. E.g.\ for simulations with 
$N=10^5$ particles a GRAPE3af (with 8 GRAPE processors) is 
competitive with a TREE--code running on a CRAY T90. 

  Simultaneous to the development of efficient stellar N-body integrators
mainly two methods (based on particles) for the treatment of a dissipative, 
star--forming gas component have been applied:
The {\it smoothed particle hydrodynamics} scheme (SPH)
solves the gasdynamical equations, i.e.\ the 
diffuse nature of the interstellar medium (ISM) is
emphasized (e.g.\ Hernquist \& Katz, 1989). An alternative approach focussed
on the clumpiness of the clouds in the ISM. These clouds were treated as
{\it sticky particles}: Without physical collisions with other clouds or 
close encounters they move 
on ballistic orbits like stars. However, in case of a collision
the clouds might merge or loose kinetic energy depending on
the adopted microphysics (e.g.\ Casoli \& Combes, 1982;
Palou\v s et al., 1993; Theis \& Hensler, 1993).

By these technical 
and numerical improvements of the last 10 years reliable N-body simulations 
even for the infall of small satellites into larger galaxies have 
become possible (e.g.\ Quinn et al., 1993; Walker et al., 1996; 
Huang \& Carlberg, 1997; Vel\'azquez \& White, 1998). 
The interest in these minor mergers raised
especially after the discovery of the Sagittarius dwarf galaxy SgrI
which seems to be an ongoing merger with the Milky Way (Ibata et al., 1994),
i.e.\ minor mergers seem to be a common phenomenon.
This is in agreement with estimates based on the cosmological CDM model 
which predicts that 85\% of the finally existing galaxies have undergone 
a merger with a companion of at least 20\% of its own mass 
(Frenk et al., 1988). Despite some debate on this high merging
rate (T\'oth \& Ostriker, 1992) tidal interaction or at least minor
merging seem to be a normal evolutionary process for a large fraction of giant
galaxies. 
 
  Though there is a large number of papers dealing with principle effects
of interactions like tidal tails or induced star formation, there
exist considerably fewer papers about the modeling of special
objects (e.g.\ Salo \& Laurikainen (1993) on NGC7752/53 or
Thomasson \& Donner (1993) on M81 and NGC3077). 
This deficiency is mainly caused by two reasons: First, for a 
detailed confrontation of theory with observation high resolution data
in configuration and velocity space are required. These should cover a large 
fraction of the space between the interacting galaxies. In principle, 
HI would be a good tracer for the global dynamics, however, there are just
a few observational sites which give data of sufficient quality. 

 The second
problem is the large parameter space 
resulting in two connected difficulties: finding a good fit at all and
determining its uniqueness (or other acceptable parameter sets).
Observationally only three kinematical quantities -- the projected position 
on the sky and the line-of-sight velocity -- can be measured. So, for
each galaxy the line-of-sight position and the two components of the
velocity in the plane of the sky are unknown. Another parameter,
the galactic mass, depends on the availability of velocity data, the 
determination of the distance, and the reliability of the conversion from
velocity to masses. These 14 parameters just fix the orbit in case
of a two--body interaction. Moreover, one has to specify the
parameters that characterize both stellar systems, e.g.\ characteristic
scales, orientation, or rotation of a disc. By this one ends up with 
a high--dimensional parameter space which is -- even including some
'prejudices' about the system -- in general too large for
a standard search method. For instance, let us assume that the orientation,
rotation, and scale length of the discs, the galactic masses and all the 
6 available kinematic data are known. The remaining parameter space has
still 6 dimensions. A regular grid with a poor coverage of 10
grid points per dimension demands $10^6$ models or 340 years GRAPE
simulation time (assuming 3 CPU--hours for a complete calculation) or
still a month with a restricted N-body program (assuming 3 CPU-seconds
per simulation). 

An alternative approach to investigate large
parameter spaces are genetic algorithms (Holland, 1975; Goldberg, 1989). The
basic idea is to apply an evolutionary mechanism including 'sexual'
reproduction operating on a set of parameters which are selected for
parenthood according to their 'fitness'. The fitness is determined
by the ability of the parameters to reproduce the observations,
whereas the reproduction is done by 'cross--over' of the parental
parameters and applying a 'mutation' operator. Though genetic algorithms (GA)
have been used in many branches of science, there are just a few
applications in astrophysics, e.g.\ for fitting rotation curves or
analysis of Doppler velocities in $\delta$ Scuti stars (for a review see 
Charbonneau, 1995). Recently, Wahde (1998) demonstrated the ability
of genetic algorithms to recover orbital parameters for artificial
data generated by N-body simulations of strongly interacting galaxies. 
He stressed that in these systems the positional information can be
already sufficient to determine the orbital elements.

   In this paper I want to demonstrate how different N-body techniques
can be combined in order to create a model for the dynamics of interacting
galaxies provided the intensity distribution is available. 
The scheme consists basically of two
steps. In the first step restricted N-body simulations will be performed
to constrain the model parameters: If a sufficient accurate data set is
available, the genetic algorithm will be used to find the orbital
parameters and to check the uniqueness of the result. Additionally,
the sensitivity of the solution can be checked by an extended parameter 
study. Alternatively, in case of insufficient data, 
one could at least use the restricted N-body models to get 
a first guess of the parameters, by this creating an artificial 
intensity map and then applying the genetic algorithm to check its uniqueness.
In the second step self--consistent models are necessary in order
to 'tune' the parameters and to check the applicability of the restricted 
N-body method. These checks have to address two questions: Is the 
neglect of self--gravity acceptable? Does gas dynamics alter 
the results significantly? Technically this means that self--consistent
simulations with and without gas have to be performed. 

  The numerical models of this paper are performed with respect to NGC4449
which is a prototype of an actively star--forming galaxy similar
in size and mass to the Large Magellanic Cloud. NGC4449
is surrounded by an extended HI halo which shows some distortion
(van Woerden et al., 1975; Bajaja et al., 1994). Hunter et al.\ (1998)
published recently high resolution HI images taken at the VLA showing a 
large scale lopsided structure. Though there seems to be
a close dwarf companion, DDO 125, it is not clear whether the observed
streamers are tidal tails caused by an interaction with DDO125
(Hunter et al., 1998). For example, tidal distortions 
are missing in the optical 
image of both galaxies and no clear bridge has been detected. 
However, assuming a face--on encounter Theis \& Kohle (1998) demonstrated 
qualitatively by N-body simulations that the main
HI--features could stem from such an encounter provided the mass
of DDO125 exceeds 10\% of the mass of NGC4449. An alternative 
scenario assumes that the HI structure is an ongoing infall
of gas onto NGC4449. This might be caused by the encounter
with DDO125 followed by a compression and cooling of the gas resulting in
an infall of the gas (Silk et al., 1987). However, it seems to be 
difficult that the observed large--scale, lopsided and regular structure 
in NGC4449 can be maintained over a typical timescale like the 
orbital period which is longer than 1 Gyr for the outer streamer. Already
the high internal velocity dispersion of 10 km\,s$^{-1}$ would destroy
the streamers on a timescale of $\nat{2}{8}$ yr (Hunter et al., 1998).
In this paper I investigate the interaction scenario addressing the 
question whether the observed structures can be formed by tidal
interaction at all and what the constraints on the mass distribution
or the orbits of both galaxies are.

Sect.\ \ref{observations} summarizes the observational data of NGC4449 
and DDO125 relevant for the models of this paper. The different
modeling techniques and the numerical results are described
in Sect.\ \ref{numerics}: First, the restricted N-body
method is introduced (Sect.\ \ref{restricted}) and a set of
first results is shown in Sect.\ \ref{firstguess} and \ref{parameterstudy}.
Afterwards the genetic algorithm approach is explained in Sect.\ 
\ref{geneticalgorithm} and its results are shown in Sect.\ \ref{unique}.
Finally, the restricted N-body simulations are compared with 
self-consistent models with (Sect.\ \ref{stickyparticles}) and 
without gas (Sect.\ \ref{scnogas}). The results of these different 
approaches are summarized and discussed in Sect.\ \ref{discussion}.
%
\section{The case of NGC4449 and DDO125}
\label{observations}
%
\subsection{NGC4449}

   When the optical images of NGC4449 and the Large Magellanic
Cloud are compared, many similarities like the presence of 
a bar, the size or the total luminosity are found. 
However, different to the LMC, which has the Milky Way in its neighbourhood,
NGC4449 seems to be a perfect candidate to study the properties of a 
genuine Magellanic type irregular unaffected by the tidal field of a 
large galaxy. Its distance of
2.9 -- 5.6 Mpc\footnote{Most of the distance estimates of NGC4449 differ 
mainly because of the assumed value $H_0$ for the Hubble constant. 
Independently of $H_0$ Karachentsev \& Drozdovsky (1998) 
derived from the photometry of the 
brightest blue stars a distance of 2.9 Mpc. However, the error of this
measurement is unclear. In accordance with Hunter 
et al.\ (1998) a distance of 3.9 Mpc will be assumed in this paper.
Thus, $1^\prime$ corresponds to 1.1 kpc.} 
allows also
for detailed investigations which were performed in the
visual (e.g.\ Crillon \& Monnet, 1969; Hunter \& Gallagher, 1997),
in the radio band (e.g.\ van Woerden et al., 1975; Klein et al., 1996), in
the HI emission line (e.g.\ Bajaja et al., 1994; Hunter \& Gallagher, 1997;
Hunter et al., 1998), in the CO emission line (e.g.\ Sasaki et al., 1990; 
Hunter \& Thronson, 1996; Kohle et al., 1998), 
in the infrared (e.g.\ Hunter et al., 1986), 
in the FUV (e.g.\ Hill et al., 1994) and in
X--rays (e.g.\ della Ceca et al., 1997; Bomans \& Chu, 1997).
Inside the Holmberg diameter of 11 kpc
NGC4449 shows ongoing strong star formation along a bar-like 
structure (Hill et al., 1994). This activity seems to induce a 
variety of morphological features like filaments, arcs, or loops which have
a size of several kpc (Hunter \& Gallagher, 1997).

\begin{figure}[t]
     \caption[ ]{\it
       Contour map of the integrated HI--image of NGC4449. NGC4449 is
       at the center of the image and DDO125 is located at the
       almost circular contours in the South. This image
       was kindly provided by Deidre Hunter.}
     \label{hunterobservation}
\end{figure}

Already the early radio observations
of van Woerden et al.\ (1975) exhibited a large HI--structure
exceeding the Holmberg diameter of 11 kpc by a factor of 7.5.
Recent observations by Bajaja et al.\ (1994) with the Effelsberg telescope 
and more detailed results by Hunter et al.\ (1998) with
the VLA revealed a complex structure of several components (Fig.\ 
\ref{hunterobservation}):
An elongated ellipse of HI gas centered on NGC4449 has a total mass
of $\sim \nat{1.1}{9} \msun$. This gas rotates rigidly inside
11 kpc reaching a level of 97 km\,s$^{-1}$ in the deprojected velocity.
Outside 11 kpc the rotation curve is flat. Bajaja et al.\ (1994)
derived a dynamical mass of $\nat{7.0}{10} \msun$ inside 32 kpc and a HI 
gas mass fraction of 3.3\%. Within 11 kpc the dynamical mass
is $\nat{2.3}{10} \msun$ and the mass inside the optical radius 
of about 5.6 kpc is $\nat{3}{9} \msun$. Applying single dish data for 
a $21^\prime$ beamwidth Hunter et al.\ (1998) derive about $10^{10} \msun$.
Kinematically interesting is the counterrotation of the HI gas
inside the optical part of NGC4449 with respect to the outer regions.
For the central ellipse Hunter et al.\ (1998) determined a
position angle of $230^\circ \pm 17^\circ$ and an inclination of
$60^\circ \pm 5^\circ$.

South of the galactic center a streamlike 
structure of 25 kpc emanates which abruptly splits into two parts:
A small spur (5 kpc) which points towards DDO125, a close irregular
galaxy, and a long extended stream. 
The latter first points 24 kpc to the
north (in the following: {\it 'vertical streamer'}), 
then 49 kpc north--east and, finally, 27 kpc to the south--east
(together called the {\it 'northeastern streamer'}),
by this covering 180$^\circ$ around NGC4449. Remarkable are the long
straight sections and the abrupt changes of direction. 
The total mass of the streamers measured by the VLA observations
is about $\nat{1}{9} \msun$. Hunter et al.\ (1998) pointed
out that by comparison with Effelsberg data there is an additional 
diffuse HI mass not detected in the VLA observations which amounts to
two--thirds of the HI in the extended structures.

\subsection{DDO125}

   DDO125 is located in the south--east of NGC4449 at a projected
distance of 41 kpc. The line-of-sight velocity difference between DDO125
and NGC4449 is only about 10 km\,s$^{-1}$. Tully et al.\ (1978) found
rigid rotation inside the optical radius of DDO125 and an inclination of the
adopted disc of $50^\circ$. From this they derived a total mass of 
$\nat{5}{8} \msun$ (corrected for a distance of 3.9 Mpc) which is
in agreement with single dish observations ($21^\prime$ beamwidth) 
of Fisher \& Tully (1981). However, Ebneter et al.\ (1987) reported 
VLA--observations which show a linear
increase of the rotation curve out to the largest radii where HI
has been detected, i.e.\ out to twice the optical radius. This already
allows for an estimate of a lower mass limit for DDO125 
which is a factor 8 larger than
the value of Tully et al.\ (1978), i.e.\ $\nat{4}{9} \msun$. Since
neither flattening of the rotation curve nor any decline has been observed,
this value is only a lower limit. If one compares that mass with the 
{\it total} mass of NGC4449 derived by Bajaja et al.\ (1994) one gets a mass
ratio of about $q \approx 5.7\%$ which is, however, just a lower limit. 
Comparing the masses inside the fiducial range of rigid rotation gives 
a higher mass ratio of $q \approx 17.4\%$.

%
\section{Numerical Modeling}
\label{numerics}

For the simulations described in this paper three different N-body 
approaches have been applied: The method of restricted N-body simulations
(Pfleiderer \& Siedentopf, 1961; Toomre \& Toomre, 1972) is described
in Sect.\ \ref{restricted}. The self--gravitating models are 
split into calculations without gas (Sect.\ \ref{scnogas}) and simulations
which include gas by means of sticky particles (Sect.\ \ref{stickyparticles}).

  The advantage of the first method is the reduction of the full N-body
problem to $N$ single body problems and the use of particles as 
'test particles' which gives high spatial resolution at the
places of interest for almost no computational cost. Furthermore,
the CPU--time scales linearily with particle number which is
superior to almost all N-body codes. Because of this
the calculation is 1000 times faster than
self--consistent simulations. Hence, the restricted N-body calculation
is at the moment the only method which allows about $10^4$
simulations in a couple of CPU-hours. Such a fast computation is a 
prerequisite for all systematic searches in parameter space like the 
genetic algorithm described in Sect.\ \ref{geneticalgorithm}.

On the other hand, the simplified treatment of the galactic potential in 
restricted N-body simulations prohibits
any effect of the tidal interaction on the orbits of both
galaxies, e.g.\ no transfer of orbital angular momentum into galactic spin
is possible (i.e. no merging). In order to overcome this 
problem, the results of restricted N-body models and genetic algorithms
must be compared with detailed self--consistent simulations. In a first
step the stellar dynamical applicability of the restricted models is
checked by a comparison with self--gravitating systems calculated with
a GRAPE3af special purpose computer (Sugimoto et al.\ 1990; see
Sect.\ \ref{scnogas}). 

   So far the dynamics of the system has been investigated in terms of
stellar dynamics, though in many cases (and also in the
case of NGC4449) HI {\it gas} is observed. In general, the effects of 
neglecting gasdynamics are not clear. E.g.\ a purely stellar dynamical 
ansatz was successfully applied for the system M81 and NGC3077 
(Thomasson \& Donner, 1993), whereas in other systems the dynamics of the gas
can deviate strongly
from the behaviour of the stars (e.g.\ Noguchi, 1988; Barnes \& Hernquist,
1996). Therefore, one has to compare stellar and gas dynamical results
which is done in this paper by a sticky particle method (Sect.\ 
\ref{stickyparticles}).\\

The units are chosen to 1 kpc, $\nat{7}{10} \msun$ (the fiducial
mass of NGC4449) and G=1, which result
in a time unit of 1.78 Myr and a velocity unit of 549 km\,s$^{-1}$.

%
\subsection{Restricted N-body models}
\label{restricted}

  The basic idea of restricted N-body simulations is to
reduce the N-body problem to $N$ 1--body problems by assuming that
the gravitational potential is given by a simple relation, e.g.\ a
two-- or a few--body problem which has a known 
\mbox{(semi-)} 
analytical solution or can be solved by fast standard methods. Here
I assume that the gravitational forces on each particle are given
by the superposition of the forces exerted by two point--like objects
(the galaxies) moving on Keplerian orbits. Thus, for the orbits one has to
specify 12 parameters which reduce to 6 in the center-of-mass frame: 
The orbital plane is fixed by the inclination angle and the argument of 
the ascending node. The orbit itself is characterized by its
eccentricity and the location
of the pericenter, i.e.\ the pericentric distance and the argument
of the pericenter. Finally, the initial location of the
'reduced' particle (or the time of pericentric passage) has to be
specified to fix the phase. For a comparison with the observations 
the masses of the galaxies and the phase of the observation have to be 
supplied, too. Three of the six orbital parameters can be fixed
by the observed location of both galaxies on the plane of sky 
($x$-$y$ plane) and the line-of-sight velocity. The masses 
can be estimated by the velocity profiles of the galaxies provided 
the distance to them is known. 

In this paper the time of pericentric passage is set to zero. As input
the orbital eccentricity and inclination and the minimum distance are
given. Furthermore, the directly
observable parameters, i.e.\ the masses of the galaxies, the projection of
the relative position of both galaxies on the plane of sky, and their
line-of-sight velocity difference are supplied. They are used to
determine the argument of the pericenter, the phase (or time since
pericentric passage) of the observed system and the argument of
the ascending node of the orbital plane. Additionally, the initial distance
or phase has to be specified: In case of parabolic and hyperbolic
encounters the distance was set to 100 kpc, which guarantees a sufficient
low tidal influence at the start of the simulation. In case of an elliptical
encounter, the situation is more difficult, because the apocentric distance 
might be small, especially if the eccentricity is low. Thus, the system 
might be affected permanently by tidal forces or even a series of 
repeated encounters which make the usage of the restricted N-body 
method doubtful. For these elliptical orbits the initial azimuthal angle of 
the orbit was set to 120$^\circ$ (0$^\circ$ corresponds to pericenter).

  The test particles are arranged in a flat disc moving on circular
orbits. The disc itself is characterized by its orientation, i.e.\
its inclination and position angle, the scalelength and the outer edge.
Additionally, an inner edge can be specified in order to avoid a waste
of computational time by integrating tidally unaffected orbits well
inside the tidal radius of the galaxy.
  In most of the restricted N-body simulations the test particles are 
distributed in 30 to 60 equidistant rings of 125 particles, i.e.\ a total
of $N=3750$ to $7500$ particles. The typical spacing
of the rings is 0.7 to 1.5 kpc depending on the outer edge of the disc.
The time integration of the equations of motion 
is performed by a Cash--Karp Runge--Kutta integrator of fifth 
order with adaptive timestep control (Press et al., 1992). In order
to prevent a vanishingly small timestep by an accidental infall of a test
particle to the galactic center, the gravitational force is
softened on a length scale of 1 kpc. In order to speed-up
the calculation the galactic positions are tabulated by cubic splines
and the integrator uses these look-up tables. On a SPARC10 with a 
150 MHz CPU a typical 7500-particle encounter takes 23 CPU seconds,
which can be reduced by a factor of 2-3 by applying an inner edge
of 10 kpc. For a typical genetic algorithm simulation with 900 particles 
and no offset for an inner edge 2 CPU seconds are required per simulation.

%
%
%
%
\subsection{A first guess}
\label{firstguess}

\begin{figure}
   \caption[ ]{\it Temporal evolution of the particle positions projected
     onto the plane-of-sky for different times. Closest approach is at 
     $t=0$. The final diagram corresponds to a projected distance of 
     41 kpc. The mass ratio of both galaxies is set to $q=0.2$. The original
     HI--disc has an inclination of $i=60^\circ$, a position angle
     $\alpha = 230^\circ$ and a radial extent of $\rmax=40$ kpc. The particles
     inside a radius of 10 kpc are omitted. The contour lines in the lower
     right diagram show the normalized surface density distribution 
     corresponding to the
     particle distribution in the lower left diagram. One time unit
     corresponds to 1.78 Myr.
      }
     \label{resevol}
\end{figure}

   In a series of simulations the orbital parameters
(minimum distance $\rmin$, the eccentricity $\epsilon$, the mass
ratio $q$ and the orientation of orbital plane) as well as the disc parameters 
(size $\rmax$ of the disc, its position angle $\alpha$ and inclination $i$) 
are varied (see Table \ref{modeltable}). 

The temporal evolution of the primary galaxy (NGC4449) in the reference model A 
is displayed in Fig.\ \ref{resevol}. At the beginning both galaxies
have a distance of 100 kpc corresponding to 995 Myr before closest
approach. The particles inside 10 kpc are omitted for an easier 
identification of the evolving structure. This neglect does not
alter the results, since the related particles are only 
weakly affected by the interaction, as the persistence of the
elliptical shape of the 'central hole' demontrates. At the
moment of closest approach ($t=0$), only the outer region of
NGC4449's disc reacted on the intruder. 71 Myr later
(i.e.\ system time 40),
a dense outer rim starting at the projected position of
DDO125 and pointing to the north begins to form. Secondly,
a weak bridge-like feature seems to connect both galaxies.
Both features are more pronounced after 178 Myr.
At that time an additional tidal feature commences to evolve
north of the center. When DDO125 has a projected distance of 41 kpc which
occurs 362 Myr after closest approach, three features are discernible:
The remnant of the former 'bridge' starts south of the center of
NGC4449 pointing to the south-west. The 'vertical feature' starts
at the end of the first structure and points to the north.
The third structure is a long tidal arm which starts west of NGC4449
pointing to the north-east before turning again back to the south-east.

A comparison of the intensity map of model A (lower right diagram in
Fig.\ \ref{resevol}) with the observations (Fig.\ \ref{hunterobservation}) 
demonstrates that the main observational features
can be reproduced qualitatively by the reference model A, i.e.\ a
parabolic orbit, a mass ratio $q=0.2$ and a minimum distance 
$\rmin=25$ kpc.
The parameters for the disc orientation are chosen in agreement 
with Hunter et al.'s suggestion of $i=60^\circ$, $\alpha=230^\circ$,
whereas the disc radius was set to $\rmax=40$ kpc.
Though this numerical model is not a detailed fit of the data, 
the characteristic sizes and locations of the streamers as well 
as the unaffected disc can be found. The material located between the 
streamer and the disc, which is not seen in Hunter's data, could
correspond to the extended HI-mass detected in the single dish 
observations by van Woerden et al.\ (1975). In the following sections 
I will discuss the influence of the individual parameters. The global
uniqueness of the reference model is investigated in
Sect.\ \ref{unique}.

%
\subsection{Parameter study}
\label{parameterstudy}

  In this section the influence of the basic parameters, 
i.e.\ the orientation of the orbital plane, the orbital eccentricity
and the minimum distance, the mass ratio of both galaxies, the orientation
of the HI disc and its extent are discussed (s.\ Table \ref{modeltable}).
In the following simulations all parameters except the one of interest
are kept constant with respect to model A. Thus, the vicinity of the
reference model in parameter space is studied.\\

\begin{table}
  \label{modeltable}
  \centerline{\Large \it Model parameters of the restricted N-body simulations}

  \vspace*{0.1cm}

  \begin{center}
  \begin{tabular}{|l || c|c|c|c || c|c|c || l|}
    \hline
    model & $\epsilon$ & $q$  & $d_{\rm min}$ & $i$ & $i_d$ & $\alpha_d$ & $R_{\rm max}$ & comment \\ \hline\hline
     A    &  1.0       & 0.2  &  25           & 40  &   60  & 230        &  40  & reference model\\ \hline\hline

     B1   &  0.5       & 0.2  &  25           & 40  &   60  & 230        &  40  & eccentricity \\ \hline
     B2   &  0.8       & 0.2  &  25           & 40  &   60  & 230        &  40  & \\ \hline
     B3   &  1.5       & 0.2  &  25           & 40  &   60  & 230        &  40  & \\ \hline
     B4   &  5.0       & 0.2  &  25           & 40  &   60  & 230        &  40  & \\ \hline\hline

     C1   &  1.0       & 0.01 &  25           & 40  &   60  & 230        &  40  & mass ratio \\ \hline
     C2   &  1.0       & 0.05 &  25           & 40  &   60  & 230        &  40  & \\ \hline
     C3   &  1.0       & 0.1  &  25           & 40  &   60  & 230        &  40  & \\ \hline
     C4   &  1.0       & 0.3  &  25           & 40  &   60  & 230        &  40  & \\ \hline
     C5   &  1.0       & 0.4  &  25           & 40  &   60  & 230        &  40  & \\ \hline\hline

     D1   &  1.0       & 0.2  &  15           & 40  &   60  & 230        &  40  & minimum distance \\ \hline
     D2   &  1.0       & 0.2  &  20           & 40  &   60  & 230        &  40  & \\ \hline
     D3   &  1.0       & 0.2  &  30           & 40  &   60  & 230        &  40  & \\ \hline
     D4   &  1.0       & 0.2  &  35           & 40  &   60  & 230        &  40  & \\ \hline
     D5   &  1.0       & 0.2  &  40           & 40  &   60  & 230        &  40  & \\ \hline\hline

     E1   &  1.0       & 0.2  &  25           & 10  &   60  & 230        &  40  & orbital inclination \\ \hline
     E2   &  1.0       & 0.2  &  25           & 20  &   60  & 230        &  40  & \\ \hline
     E3   &  1.0       & 0.2  &  25           & 30  &   60  & 230        &  40  & \\ \hline
     E4   &  1.0       & 0.2  &  25           & 50  &   60  & 230        &  40  & \\ \hline
     E5   &  1.0       & 0.2  &  25           & 70  &   60  & 230        &  40  & \\ \hline\hline

     F1   &  1.0       & 0.2  &  25           & 40  &   40  & 230        &  40  & disc inclination \\ \hline
     F2   &  1.0       & 0.2  &  25           & 40  &   50  & 230        &  40  & \\ \hline
     F3   &  1.0       & 0.2  &  25           & 40  &   70  & 230        &  40  & \\ \hline\hline

     G1   &  1.0       & 0.2  &  25           & 40  &   60  & 210        &  40  & pos.\ angle of disc \\ \hline
     G2   &  1.0       & 0.2  &  25           & 40  &   60  & 250        &  40  & \\ \hline\hline

     H1   &  1.0       & 0.2  &  25           & 40  &   60  & 230        &  25  & outer edge of disc \\ \hline
     H2   &  1.0       & 0.2  &  25           & 40  &   60  & 230        &  30  & \\ \hline
     H3   &  1.0       & 0.2  &  25           & 40  &   60  & 230        &  35  & \\ \hline
     H4   &  1.0       & 0.2  &  25           & 40  &   60  & 230        &  50  & \\ \hline
  \end{tabular}
  \end{center}
  \caption[]{\it Parameters of the different restricted N-body models: 
     model name \mbox{(col.\ 1)}, orbital eccentricity
   $\epsilon$ (col.\ 2), mass ratio $q$ of both galaxies 
    (col.\ 3), minimum separation $\rmin$ of galactic centers
    (in kpc, col.\ 4), inclination $i$ of orbital plane 
    (in degrees, col.\ 5), orientation of NGC4449 disc (in degrees):
    inclination $i_d$ (col.\ 6) and position angle $\alpha_d$ (col.\ 7), 
    maximum radius $R_{\rm max}$ of gaseous disc \mbox{(in kpc, col.\ 8)}.}
\end{table}

\begin{figure}
   \caption[ ]{\it Dependence on the eccentricity.
     Projection of the particle positions onto the plane-of-sky in the 
     moment of a projected distance of 41 kpc. The eccentricities of the models
     are 0.5 (upper left, model B1), 0.8 (upper right, B2),
     1.0 (A, reference model), 1.5 (B3) and 5.0 (B4).
      }
     \label{resecc}
\end{figure}

\noindent
{\bf Eccentricity.} The eccentricity mainly influences the interaction
timescale which is defined here as the time between maximum approach and
a projected distance of 41 kpc, i.e.\ the moment of observation. 
An eccentricity of 5.0 (corresponding to a hyperbolic encounter)
decreases the interaction time by
a factor 2.6 resulting in less 'damage' of the primary galaxy
(Fig.\ \ref{resecc}): Only the remnant of the 'bridge' feature can be 
discerned, whereas the other two streamers found in the reference model
are missing completely. With decreasing eccentricity the north-eastern
streamer becomes pronounced before the vertical structure arises. The latter
is weakly expressed for $\epsilon=1.5$. Reducing $\epsilon$ further
to bound orbits makes the features even more prominent. In case
of $\epsilon=0.5$ the north-eastern and the vertical streamer form
even a single structure clearly detached from the center. However,
due to the strength of the interaction a second broad tidal 'arm'
is ejected from NGC4449 leaving a low-density region south to the 
northern tip of the north-eastern streamer.\\

\noindent
{\bf Mass ratio.} The variation of the mass ratio gives a lower limit of 
  about 10\% 
  in order to create a sufficient tidal response (Fig.\ \ref{resmass}). 
  If the mass ratio exceeds 0.3, the 'vertical feature' becomes less vertical
  and the north-eastern streamer is more diffuse, both due to the enhanced 
  tidal pull by the intruder.\\

\begin{figure}
   \caption[ ]{\it Dependence on the mass ratio.
     Projection of the particle positions onto the plane-of-sky in the 
     moment of a projected distance of 41 kpc. The mass ratios are
     - starting with the upper left diagram - 0.01 (model C1), 0.05 (C2),
     0.1 (C3), 0.2 (A, reference model), 0.3 (C4) and 0.4 (C5).
      }
     \label{resmass}
\end{figure}

\noindent
{\bf Minimum distance.} Though the interaction time is not strongly
  influenced by the minimum distance $\rmin$, it determines the maximum
  strength of the tidal field (Fig.\ \ref{resrmin}):
  For small separations the structures become more
  diffuse and the primary is much stronger affected.
  The vertical feature is much more 
  extended creating an additional tidal 'arm'
  (similar to the model with an orbital eccentricity $\epsilon=0.5$).
  If the distance of closest approach exceeds 30 kpc, almost no streamers
  are formed.\\

\begin{figure}
   \caption[ ]{\it Dependence on the minimum distance during the encounter.
     Projection of the particle positions onto the plane-of-sky in the 
     moment of a projected distance of 41 kpc. The minimum distances are
     - starting with the upper left diagram - 15 kpc (model D1), 20 kpc (D2),
     25 kpc (A, reference model), 30 kpc (D3), 35 kpc (D4) and 40 kpc (D5).
      }
     \label{resrmin}
\end{figure}

\noindent
{\bf Orbital plane.} The orientation of the orbital plane characterizes
  basically the observer's location or the line-of-sight which is fixed 
  by two parameters. For the models here only the orbital inclination
  is a free parameter, since the second parameter, the 
  position angle of the orbital plane, is fixed by the
  observed relative line-of-sight velocity of both galaxies.
  The models show that the inclination mainly affects
  the orientation of the vertical streamer (Fig.\ \ref{resiorb}):
  An inclination outside the range $[30^\circ,50^\circ]$ gives too
  much deviation from the south-north direction of the vertical
  feature.\\

\begin{figure}
   \caption[ ]{\it Dependence on the inclination angle of the orbital plane.
     Projection of the particle positions onto the plane-of-sky in the 
     moment of a projected distance of 41 kpc. The inclinations are
     10$^\circ$ (upper left, model E1), 20$^\circ$ (upper right, E2), 
     30$^\circ$ (E3), 40$^\circ$ (A, reference model), 50$^\circ$
     (E4) and 70$^\circ$ (E5).
      }
     \label{resiorb}
\end{figure}

\begin{figure}
   \caption[ ]{\it 
     Dependence on the extension of the HI distribution in NGC4449.
     Projection of the particle positions onto the plane-of-sky in the 
     moment of a projected distance of 41 kpc. The radial extensions are
     25 kpc (upper left, model H1), 30 kpc (upper right, H2),
     35 kpc (H3), 40 kpc (A, reference model) and 50 kpc (H4).
      }
     \label{resrhig}
\end{figure}

\begin{figure}
   \caption[ ]{\it Dependence on the orientation of the HI disc in NGC4449.
     Projection of the particle positions onto the plane-of-sky in the 
     moment of a projected distance of 41 kpc. The orientation is characterized
     by the inclination $i_d$ of the disc and the position angle $\alpha_d$:
     $i_d = 60^\circ, \alpha_d = 230^\circ$ 
     (upper left, model A, reference model); 60$^\circ$, 210$^\circ$ 
     (upper right, G1); 
     60$^\circ$, 250$^\circ$ (G2); 40$^\circ$, 230$^\circ$ (F1);
     50$^\circ$, 230$^\circ$ (F2) and 70$^\circ$, 230$^\circ$
     (F3).
      }
     \label{resdisc}
\end{figure}

\noindent
{\bf HI distribution in NGC4449.} A specification of the mass-
and especially the HI-distribution is a more difficult task, because contrary
to the well-known two-body problem the number of free parameters
characterizing the structure and internal dynamics of the interacting galaxies
is unknown and a matter of investigation itself. For example, it is
not a priori clear whether the HI is distributed like a disc, a
spheroid, or something else. 
In order to reduce the number of free parameters I will
assume here that the observed inner elongated ellipse is the remnant
of an original exponential HI disc, which can be specified by its 
orientation (which is observed!), its scale length and radial extent. 
Since the particles are treated as test particles, a variation of the
scale length does not change the final location of a particle, but
its mass which is required for the calculation of the intensity map.
A small scale length (e.g.\ 5 kpc) would confine
too much HI in the center that is unaffected by the encounter. Since
the observations show that a large fraction of HI is outside 20 kpc,
I have chosen a large scale length of 30 kpc which gives only a
weak radial decline of the HI surface density and, by this, a sufficient
amount of HI in the outer regions.

\noindent
{\sc Disc size.} 
The influence of the maximum radius $\rmax$ of the disc is 
shown in Fig.\ \ref{resrhig}. If $\rmax$ is smaller than 35 kpc, the
vertical streamer has not been formed at all indicating that this feature
stems from the HI gas far outside the center. The north-eastern
arm and the 'bridge', however, are already built up for $\rmax=25$ kpc.
With increasing disc size the vertical streamer becomes more
pronounced and longer which excludes disc sizes larger than 45 kpc.

\noindent
{\sc Disc orientation.} 
The orientation of the HI disc also strongly affects the appearance
of the primary galaxy. A variation of the position angle $\alpha$ 
by 20$^\circ$ gives already HI distributions which are incompatible 
with the observations (Fig.\ \ref{resdisc}). 
If $\alpha$ is decreased, the vertical streamer
is more extended and diffuse. On the other hand, if $\alpha$ is increased,
the north-eastern arm becomes too long and the vertical feature vanishes 
almost completely. 
Similarly, the results depend strongly on the inclination $i$:
Inclinations below $50^\circ$ give only weak bridges and weak vertical 
streamers, whereas the orientation of the 'vertical' streamers deviates 
strongly from the
northern direction for inclinations exceeding 70$^\circ$.\\

  The performed parameter study demonstrates that even 
close to the parameters of model A, the final particle 
distribution shows significant variations which are not in
agreement with the observations. Thus, the reference model
might be a good and locally unique candidate describing the 
dynamics of NGC4449 and DDO125. 
The confidence in this encounter scenario is especially enhanced
by the strong dependence on the orientation of the HI disc
of NGC4449. Qualitatively good models are only found, if the disc orientation 
agrees with the values determined independently by observations.
On the other hand, the former parameter
study only investigates a small fraction of parameter space
and, thus, the uniqueness of the solution is not confirmed.
Moreover, in general it is very tedious to find a good
model 'by hand', i.e.\ by an unsystematic non--automated search
in parameter space. Thus, the probability to accept the
first solution found seems to be very large (especially, if
CPU intensive simulations are performed), by this neglecting
other regions in parameter space which might also give acceptable
fits. In the next section an automatic search in parameter
space by means of a genetic algorithm is introduced. This allows in 
principle for an {\it ab initio} determination of acceptable 
parameter sets (if data sets of sufficient quality are available) or 
at least for a uniqueness test of a favoured scenario like
the reference model described above.

%
\subsection{A systematic search using a genetic algorithm}
\label{geneticalgorithm}

   The idea of applying models of organic evolution for optimization
problems in engineering dates back to the 1960s and 1970s 
(e.g.\ Rechenberg, 1965; Schwefel, 1977). Different to standard
gradient techniques for optimization which work deterministically (e.g.\ 
the downhill simplex method (Press et al., 1992)) Rechenberg's concept of
{\it 'Evolutionsstrategie'} is a probabilistic one: Starting with a more or 
less random individual ({\it 'parent'}), i.e.\ a single point in search space, 
a {\it 'child'} is generated by a random mutation of the parameter set 
characterizing the parent. The quality of both individuals with respect to 
the optimization problem (i.e.\ their {\it 'fitness'}) 
determines now the parent of the next generation. 
Repeating this process of mutation and selection
improves the quality of the individual monotonously. This simple evolutionary
strategy, which is just a special implementation of evolutionary algorithms
(EAs) in general, demonstrates the advantages of these methods: Compared to
complete grids in the parameter space the probabilistic, but oriented 
nature of the evolutionary search strategy allows for an efficient check
of a high--dimensional parameter space. Compared to gradient methods
which are very fast near the optimum, EAs do not need any gradient
information which might be computationally expensive. Additionally, they
depend only weakly on the starting point and - most important - 
they are able to leave local optima. 
The price for these features is a large number of fitness
evaluations or test points in search space before converging to a good 
solution.

   Holland (1975) extended the {\it Evolutionsstrategie} of Rechenberg
by abandoning asexual reproduction: He used a population of individuals 
instead of a single individual (and its offsprings). From the population
two individuals were selected according to their fitness to become
parents. These parents represent two points ({\it 'chromosomes'}) 
in search space
and the corresponding coordinates are treated like genes on a chromosome.
These chromosomes are subject to a {\it cross--over} and a 
{\it mutation} operator
resulting in a new individual which is a member of the next generation.
This breeding is repeated until the next generation has been formed. 
Finally, the whole process of sexual reproduction is repeated iteratively
until the individuals representing a set of points in parameter space 
confine sufficiently one or several regions of high fitness. Despite
many differences in the detailed implementations of genetic algorithms (GAs),
the basic concept of an iterated application of randomized 
processes like recombination, mutation, and selection on a population of 
individuals remains in all GAs (Goldberg, 1989; B\"ack, 1996). 

For the optimization problem of fitting an N-body simulation to a given
observation, I will briefly describe the main ingredients of my fitting
GA-program in the following sections. The GA used here is a slightly 
modified version
of P.\ Charbonneau's code {\sc pikaia} (for details see Charbonneau (1995)).

\begin{figure}[t]
   \caption[ ]{\it Fitness and best fit of an early generation during the GA
       fitting procedure. The figure displays the artificial original data
       (left column: positions of the particles in the $x$-$y$--plane
        (upper left) and the resulting intensity contour lines
        (linear spacing, lower left), the best fit of the actual generation
        (middle column) and information about the fitness history, i.e.\
        the maximum fitness (upper right) and the actual fitness distribution
        (lower right). The underlying grid in the left and middle column
        is used to determine the intensities.
           }
   \label{fig3}
\end{figure}

\subsubsection{Fitness and parent selection}

 In order to determine the fitness of an individual (i.e.\ a special parameter
set), the N-body simulation has to be performed and compared with the
observation. This is done by mapping the particles on a quadratic grid 
of size 60 kpc centered on NGC4449. The resolution
of the grid was set to 7x7 in order to allow for a better resolution of the
tidal features of an encounter. For each grid cell the 'intensity' was
calculated by summing up the masses of the individual particles in that
cell. The conversion to observed intensities can be performed by assuming
a mass--to--light ratio for the particles (or treating it as another free 
parameter) or by normalizing the intensities. Here the intensities are
normalized to the total intensity of the whole grid. The quality can be
measured now by the relative deviation $\delta$ of the intensities in both 
maps:
\begin{equation}
                \delta \equiv 
                     \sum_{i ({\rm cells})} \, \, \, 
                     \frac{|I_{{\rm ref},i} - I_{{\rm mod},i}|}
                      {\max(I_{{\rm ref},i},I_{{\rm mod},i})}
\end{equation}

The fitness $f$ is then estimated by
\begin{equation}
         f \equiv \frac{1}{1+\delta}\,. \hfill
\end{equation}

\noindent
  In case of an impossible parameter configuration (e.g.\ a circular orbit
with a distance less than the observed projected distance) or any convergence
problems the fitness of that point in search space is set to zero prohibiting
any offsprings. Typically only a few models in $10^4$ parameter sets fail,
most of them in the first few generations before the system starts to 
approach the search space region of interest.

  In order to determine the parents of the next generation the members
of the actual generation are ranked by their fitness. Then the parents are
selected randomly, whereas the probability to be chosen is proportional
to the rank ({\it 'roulette--wheel selection'}).

  A typical example for the status of the GA after a couple of generations
is shown in Fig.\ \ref{fig3}. In the left column the positions of
the original data are discretized on the grid to yield the intensities
displayed on the lower left. It should be noted that the resolution of the
grid is about 8.6 kpc which rather exceeds the thickness of the tidal 
structures. In the middle column the best fit of the actual (here 7th)
generation agrees already qualitatively with the original. The quality
of the fit is calculated only by a comparison of the lower two intensity 
diagrams. The fitness distribution in the population resembles a Gaussian
at the beginning of the fitting procedure.

\subsubsection{Coding of parameters}
\begin{center}
\begin{figure}[ptbh]
   \psfig{figure=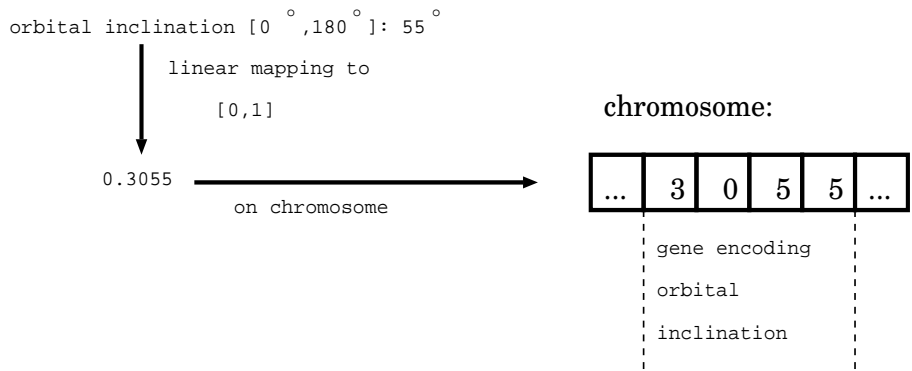,angle=270,height=5cm,width=12cm}
   \caption[ ]{\it Schematic diagram for the coding of parameters on
      a numerical chromosome.
           }
   \label{coding}
\end{figure}
\end{center}

  In organic systems the information is coded in genes in form of
sequences of the four nucleotide bases where a triplet of them (a codon) 
encodes one amino acid, the building block of proteins. 
The genes themselves are 
arranged in one or several chromosomes which contain all the genetic
information (or the genotype). The phenotype which is subject to the
selection process is the result of the genetic information and the
interaction with its environment. 

   Translated to our fitting problem, the
phenotype is the final intensity map created by a special choice of
parameters or initial conditions. The latter correspond to the proteins
which express the genetic information physically. The mapping between
the parameters and the genes or the number of chromosomes is not fixed in GAs. 
Many GAs apply a binary alphabet for encoding the parameters
(e.g.\ Holland, 1975; Goldberg, 1989). In Charbonneau's program
{\sc pikaia} a  decimal encoding is implemented. For my calculations I
combined four decimal digits giving a number between zero and one.
This 'gene' has been mapped
to a real parameter by a linear or logarithmic transformation to the
investigated parameter range, e.g.\ the orbital inclination is mapped
linearily to the range [$0^\circ$, 180$^\circ$] (Fig.\ \ref{coding}). 
All the genes are assumed to be located in linear manner on a single 
chromosome. Different to organic evolution the phenotype is completely 
determined by the genotype, no environmental influence on the individual 
is acting after its genotype is fixed.

\subsubsection{Cross--Over}

\begin{center}
\begin{figure}[ptbh]
   \psfig{figure=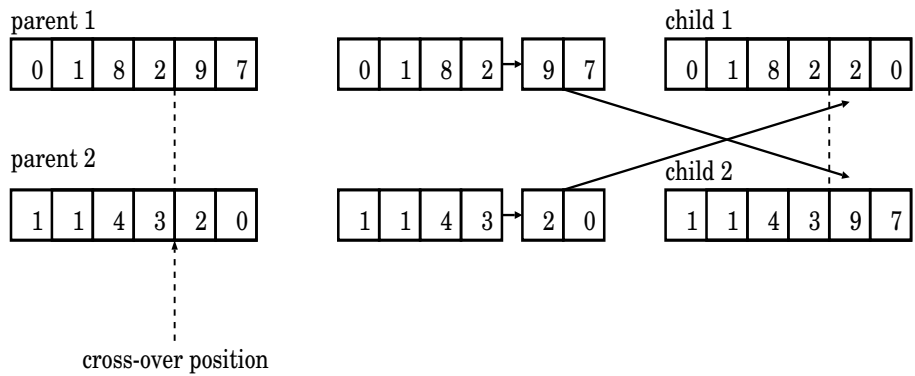,angle=270,height=5cm,width=12cm}
   \caption[ ]{\it Schematic diagram for the cross--over operation.}
   \label{crossover}
\end{figure}
\end{center}

   The main difference between the evolutionary strategies  and
genetic algorithms is the sexual reproduction of GAs. This means 
that the chromosomes of the parents are combined in a new manner by means of
a cross--over mechanism in addition to mutation. The principle idea 
is illustrated in Fig.\ \ref{crossover}: After determining the two
parents a random position on the chromosomes is selected. At this position
the chromosomes are split into two fragments which are exchanged
among the chromosomes resulting in two new ones.

\subsubsection{Mutation}
\begin{center}
\begin{figure}[ptbh]
   \psfig{figure=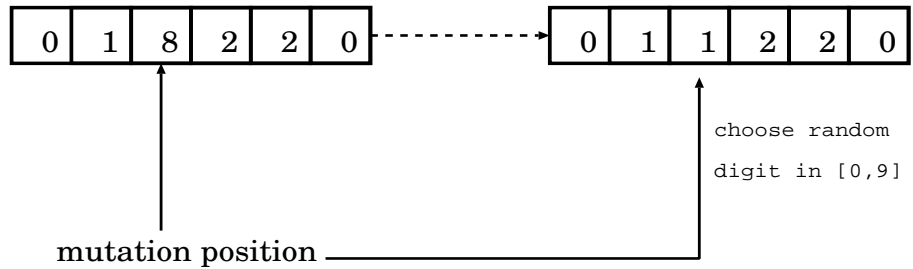,angle=270,height=3.5cm,width=12cm}
   \caption[ ]{\it Schematic diagram for the mutation process.}
   \label{mutation}
\end{figure}
\end{center}

   In order to introduce completely new information into a genepool
a mutation process, i.e.\ a random change of the information in the
chromosomes is introduced (Fig.\ \ref{mutation}). For each position
on a chromosome mutation is applied with a small probability
(typ.\ $p_{\rm mut} \approx 0.5\%$). After selecting a position for
mutation, an integer random number in the range [0,9] replaces
the original value at that position. Thus, a real change in the genetic
information takes place with a probability of 90\%, even if
that position of a gene is subject to mutation. The overall probability
for at least one 'real' mutation on a chromosome coding 6
parameters by 4 digits is about 11\%. Thus, cross--over is the most
important process introducing diversity into the population, whereas
mutation affects only each 9th member of the population. In principle,
the mutation rate can be increased, but the price would be a more
or less random search in parameter space which is already after a 
few generations inferior to the oriented search of a GA method
(Wahde, 1998). 

   A principle danger of GAs is inbreeding, i.e.\
the homogenization of the genepool by one or a few dominant 
genes. If they successfully replace other genes, cross--over does not
introduce sufficient (or -- if inbreeding is complete -- any) diversity 
into the population and thus any further improvement of the
chromosomes is prohibited. The only way to overcome this situation
is to introduce more diversity by mutation. Therefore, {\sc pikaia}
allows for a variable mutation rate which increases the mutation
probability $p_{\rm mut}$ whenever the fitness distribution becomes narrow,
i.e.\ inbreeding is suspected to operate.
On the other hand $p_{\rm mut}$ is decreased, if the fitness distribution
becomes very broad, i.e.\ the randomization is supposed to be dominant.

\subsubsection{Determination of the next generation}

   After $N_{\rm pop}$ children have been created, the old
generation is completely replaced ({\it full generational replacement})
by the next generation. The only exception is the survival of the fittest
member of the parent generation, if it is fitter than the child which 
would replace it ({\it elitism}). The latter mechanism prevents the
system from forgetting successful solutions. 

  The initial population is randomly chosen in the accessible
parameter space. A typical population size is about 
$N_{\rm pop} = 100 \dots 200$ members. The GA converges typically after
about $N_{\rm gen} = 100$ generations. Thus, $10^4$ or a few $10^4$
N-body simulations are required for a single GA fit. Self--consistent
N-body models like the mentioned 3--CPU-h GRAPE models would still need
3.4 years, whereas the restricted N-body models reduce the
CPU--requirement to 5.6 hours on a 150MHz--Sparc10.

%
\subsection{Are the results unique?}
\label{unique}

\begin{figure}[h]
   \caption[ ]{\it Best fit model during the course of a GA fitting procedure.
       The projection of the particles in $x$-$y$--plane and the corresponding 
       grid for the intensity evaluation are displayed: The original data
       (upper left), the best fit of the GA after initialization (upper middle),
       after the first breeding (upper right), after 11 generations (lower left)
       and at the end of the fitting procedure after 100 generations
       (lower middle). The evolution of the maximum fitness is shown in
       the lower right diagram. The number of test particles per simulation is 900.
           }
   \label{gaevol}
\end{figure}

\begin{figure}
   \caption[ ]{\it Development of six parameters of the best fit model during
     a GA run. The parameters are the mass of the secondary galaxy (upper 
     left), the minimum distance (upper right), the inclination (middle left)
     and position angle (middle right) of the orbital plane, and the
     inclination (lower left) and position angle (lower right) of the 
     disc of the primary galaxy. The filled squares show the parameters
     of the fitted artificial reference model.
      }
     \label{gaparams}
\end{figure}

\begin{figure}[t]
   \caption[ ]{\it Test of the capability of the GA (lower row) 
    to recover original data
    when these data (upper row) are of low quality or incomplete.
    Shown are runs with low resolution, i.e.\ a discretization on a 
    4x4 grid (left column), a bad pixel (here an omitted central pixel,
    middle column), and a missing structure (the vertical streamer,
    right column).
           }
   \label{gatest}
\end{figure}

  In order to check the uniqueness of the reference model A, several runs 
with different parameter sets encoded in genes were performed. Fig.\ 
\ref{gaevol} shows a typical run of the GA operating on a population of 
100 members. The reference model -- displayed in the upper left diagram --
as well as the GA models have been calculated with 900 test particles.
This makes the streamers less 
prominent, but still discernible. The next plots show the GA status
at different generations: At generation 1, the best model is just
created out of the randomly chosen initial parameter sets. However,
already at the second generation, i.e.\ the first application of the
reproduction operator, the best model clearly exhibits the north-eastern
streamer. At generation 11, even the weaker vertical feature and the
bridge feature show up. At the end of this GA run the particle distribution
of the original model A and the best fit of the GA are almost identical,
at least when compared by eye. The fitness of the best fit has strongly
increased from 0.04 to 0.14 where values above 0.1 indicate already
a very good fit. Most of the improvements were found during the first
50 generations, i.e.\ analyzing 5000 models. After that generation 
the homogeneity of the population led to inbreeding which strongly
prohibits the arise and spread of new 'better' parameters. However,
this is not a problem for the models here, because a very good fit to
all parameters is already found after 40 generations as the direct
comparison with the parameters of the reference model demonstrates
(Fig.\ \ref{gaparams}). The relative deviation of the derived
parameters from the original values is less than 15\% in all cases,
and for many of them even better than 5\%.

  In a series of different GA runs I varied the population size,
the number of generations or particles, the parameter combination encoded 
on chromosomes, the fitness function, and the random initial population.
In almost all of them the best fit gave an acceptable fit of 
the reference model which demonstrates the capability of GAs
to recover the parameters even for weak encounters. The models
which failed basically suffered from inbreeding or from a
fitness function which did not discriminate sufficiently between
high and low quality fits. Moreover,
in none of the GA runs a different region in parameter
space producing a good fit ($f>0.1$) has been found. {\it Thus, 
the model A -- or more exact the corresponding intensity map --
seems to result from a unique parameter set}. However, this
is only a 'motivated guess', not a mathematical proof!

   Another interesting question is how strongly do the
GA results depend on the completeness or accuracy of the
data. I studied this in two ways: First, I used 
a coarser grid (4x4) in order to mimic a lower resolution of
observational data (left column in Fig.\ \ref{gatest}).
Then the GA is only qualitatively able to
reproduce the reference model. Obviously, too much
information is lost, when the resolution becomes
too small. On the other hand, an increase of the number of cells to 10x10
did not improve the fit found on a 7x7 grid, but introduced 
more noise into the intensity maps. In the second test
I discarded one or several grid cells from the fitness
calculation (middle and right column in Fig.\ \ref{gatest}).
Omitting one cell does not affect the final fit, 
independently of the location of the cell (here the center).
However, if a whole feature like the vertical streamer
is neglected, the GA is unable to recover the original
data.

%
\subsection{Self--consistent models 'sin gas'}
\label{scnogas}

\begin{figure}
   \caption[ ]{\it Comparison of restricted N-body models (left) 
     with self--consistent simulations (right) for a parabolic encounter
     (upper row) and an eccentricity of $\epsilon=0.5$ (middle row). The
     projected distance to the perturber is 41 kpc. The lower row displays
     the normalized intensity profiles for the self--consistent
     simulations (left: $\epsilon=1$, right: $\epsilon=0.5$). Note that the surface
     plots are seen from north--west in order to emphasize the coherent
     streamer structure.
      }
     \label{resscon}
\end{figure}

\begin{figure}
   \caption[ ]{\it Self--consistent N-body simulation 'sin gas':
     Projection of the particle positions onto the plane-of-sky at different
     times: initially (upper row), at closest approach (middle row) and 
     at the projected distance of 41 kpc (lower row).
     The left column shows the disc
     particles, whereas the right column displays the halo particles.
      }
     \label{scondiskhalo}
\end{figure}

   The self--consistent models in this paper were performed by a direct
N-body integration using the GRAPE3af board in Kiel for the calculation
of the gravitational forces and potentials. The time integration was
done by a leap--frog scheme with a fixed timestep of 1.78 Myr. The GRAPE
board uses intrinsically a Plummer softening, the softening length
was chosen to be 0.2 kpc. These choices give a total energy conservation 
of better than 1\% during the whole simulation.

   Different to the restricted N-body models the set-up of the initial
particle configuration for self-gravitating systems is far from being
trivial. In principle, a stable stationary stellar system fulfills
the collisionless Boltzmann equation. Thus, a solution of the latter
could be used as a trial galactic system. From observations of the 
Milky Way it is known, however, that even if the Galaxy is axisymmetric,
three integrals of motion are required to construct a stationary
model. Unfortunately, only two are known analytically (the energy and
the $z$-component of the angular momentum). 
Additionally, even if a particle configuration
is generated from a distribution function depending on the integrals
of motion, it is unclear whether it is stable against
perturbations which might lead to e.g.\ the Toomre instability or 
the bar instability.
Thus, one has to check the stability on the timescale of interest
numerically.

  In this paper I use the models of Kuijken \& Dubinski (1995)
consisting of three components which are determined self-consistently
from the Boltzmann equation: The bulge follows a King
model and the halo a lowered Evans model. For the disc the
third integral is approximated by the $z$-energy, i.e.\ the energy 
in vertical oscillations. The advantage of Kuijken \& Dubinski's method
is the possibility to specify parameters like scale length or scale height
of the disc directly, whereas many other methods (e.g.\ Barnes, 1988) that
start not in virial equilibrium, readjust the mass distribution on a
dynamical timescale, which means less control on the structural parameters.
The long--term stability has been checked by simulations of an isolated
disc with $N=18000$ particles (8000 for the disc, 4000 for the bulge 
and 6000 for the halo). Within 1.5 Gyr the Lagrange radii at 10\% and
90\% mass vary only by 2-4\%. They show no systematic trend except for a
small expansion during the first 100 Myr which is probably caused by
gravitational softening. Due to two--body relaxation the scale height of
the disc increases by almost a factor of 2 within 1.5 Gyr. However, this
heating is not expected to alter the result of the dominant interaction 
significantly. The Fourier amplitudes $C_2$, defined by 
$C_m \equiv (\int_{\rm disc} \Sigma(R,\phi) r \, dr \, e^{-im\phi} d\phi) /
            (\int_{\rm disc} \Sigma(R,\phi) r \, dr \, d\phi))$ (with the
surface density $\Sigma$), are always less than 5\%, in agreement
with no significant bar or spiral pattern built up during the simulation.

   Fig.\ \ref{resscon} shows a comparison of the restricted N-body 
simulations with the self-consistent models. Except for the omitted
central region in the former and the more diffuse structure in the
latter, the large scale features are almost identical in both simulations.
The differences close to the central 10 kpc (especially the phase
of the 'bridge' feature in the $\epsilon=0.5$ model) are 
due to the extended dark halo in the self-consistent models. It is
remarkable that the vertical and the north-eastern streamer are 
not affected by the halo. This confirms the applicability of the
restricted N-body method here. In agreement with these simpler
models the intensity maps of both self-consistent models 
($\epsilon = 1.0, 0.5$) show clearly the streamer structure
(Fig.\ \ref{resscon}).
 
   Another interesting aspect concerns the response of the halo
to the intruder. Different to the disc, the halo remains almost 
unchanged by the encounter (Fig.\ \ref{scondiskhalo}). This
constancy of the halo potential would in principle allow to 
include also a rigid halo in the restricted N-body models and, hence,
also in the GA fitting procedure, by this extracting information
about the dark halo. 

%
\subsection{Self--consistent models 'con gas'}
\label{stickyparticles}

\begin{figure}
   \caption[ ]{\it
     Projection of the particle positions onto the plane-of-sky at different
     times: at closest approach (upper row), at the projected distance of
     41 kpc (middle row) and at the end of the simulation (lower row).
     Both models include self--gravity. The left column shows a purely
     stellar simulation, whereas the right column includes gas dynamics
     by means of sticky particles.
      }
     \label{scondissnodiss}
\end{figure}

    For the gasdynamical models I use a sticky particle scheme which 
treats each particle of the HI disc as a gaseous cloud. The radius of
these clouds is assumed to follow $R_{\rm cl} = 20 \cdot 
\sqrt{M_{\rm cl} / 10^6 \msun} \, {\rm pc}$ which is derived from
the mass--radius relation of Rivolo \& Solomon (1987) corrected
for glancing collisions ($M_{\rm cl}$ is the mass of a gas cloud). 
When the distance of two clouds falls below
the sum of their radii, merging
of the clouds is allowed, if their relative orbital angular momentum
is smaller than the maximum angular momentum a merged cloud
can acquire before break-up. If this
angular momentum criterion prevents a merger, the clouds keep their
kinematical data. Star formation is only qualitatively included by adopting
an upper mass limit of $\nat{7}{8} \msun$ for clouds. If they
exceed this limit no further merging is allowed, i.e.\ they are treated
like stars. However, the star formation criterion is not important
for the models of this paper. (For details of the cloud model or the
merging mechanism see Theis \& Hensler, 1993). This dissipative
scheme has been successfully applied to collapsing systems reproducing
many observed properties of spiral and elliptical galaxies. For the
long--term evolution of barred disc galaxies a comparison of this scheme with 
that of Palou\v s et al.\ (1993) for $\beta_r = \beta_t = 0.2$ gives 
basically the same results (Jungwiert, 1997). Thus, the chosen model
should be applicable for a reliable determination of the difference
between gas dynamical and purely stellar dynamical models.

     A comparison of gaseous models with dissipationless models
gives no significant difference (Fig.\ \ref{scondissnodiss}). Due
to the low gas densities only about 20-30\% of the clouds 
undergo inelastic collisions which are mainly occuring in the
denser central regions. Thus, the outer HI halo is only very
weakly influenced by (sticky particle) gas dynamics and the applicability
of purely stellar simulations for the HI seems to be possible.
It would be interesting to compare these models with an SPH
approach emphasizing the diffuse nature of the interstellar medium.

%
\section{Discussion}
\label{discussion}

\subsection{Interaction between NGC4449 and DDO125?}

   The previous simulations demonstrated that models based on the encounter 
scenario are able to reproduce the morphology of the three main streamers.
Furthermore, the interaction puts gas to the south-west of 
NGC4449, where a clumpy HI distribution is observed (Hunter et al., 1998).
Though its detailed patchy structure has not been recovered
by the models, an initial gas distribution less regular than the 
smooth disc-like one used for the numerical models might 
be responsible for this deviation. 

   The parameter study showed that already small variations
in the parameters give HI distributions which are not compatible
with the observations. Especially, the strong dependence of the
final structure on the initial orientation of the HI disc
confines its inclination and position angle within 10$^\circ$
to the values derived {\it independently} from the observations. 
This remarkable coincidence would occur necessarily in the
encounter scenario, whereas in the infall scenario it happens
just by chance. A second critical feature is the abrupt direction 
change of the streamers. In the infall scenario they mark
the path of the perturbed gas which 
collapses to the galactic center. However, it is hard to
imagine a realistic static potential that could create orbits 
with such strong
direction changes as e.g.\ from the vertical to the bridge
feature. Moreover, due to the large distance of these
features from the galactic center the potential should
be more or less spheroidal or already spherical, which makes
it even more difficult to produce such orbits.

  The most critical parameter for the interaction scenario is the
mass ratio $q$ of DDO125 and NGC4449. The simulations demonstrated that
a mass ratio smaller than 10\% cannot reproduce the streamer structure.
In order to determine $q$, both galactic masses have to be
known. In Bajaja et al.'s (1994) estimate for the mass of NGC4449 
they assumed that the HI is in rotational equilibrium even at a
distance of 32 kpc. However, in case of an interaction with DDO125
the velocities at large distances might be strongly affected by
the interaction and definitely not being in rotational equilibrium
as the 'ejection' of the tidal arms in a continuation of the simulations 
demonstrates. Hence the derived mass of $\nat{7}{10} \msun$ is 
at best an upper limit to the mass of NGC4449. A better guess
might be the dynamical mass inside the central gaseous ellipse
which seems to be unaffected by interaction. E.g.\ the dynamical
mass inside 11 kpc would be a factor 3 smaller than Bajaja's value.
Even more difficult than for NGC4449 is the mass determination
of DDO125. Observations by Tully et al.\ (1978) gave a dynamical
mass of $\nat{5}{8} \msun$ inside the optical radius of DDO125.
However, Ebneter et al.\ (1987) reported a linear increase
of the rotation curve exceeding the range observed by Tully et al.
by a factor of two. Moreover, they found no hint for a flattening
or turn-over in DDO125's rotation curve. This gives a lower mass 
limit of $\nat{4}{9} \msun$. Combining the mass estimates yield
a lower limit of the mass ratio of 6\% when using Bajaja's mass 
determination of NGC4449. A more realistic mass ratio might be
derived from the ranges of rigid rotation resulting
in $q\approx18\%$. Thus, observationally it cannot be excluded that
the mass ratio is below the critical one for the interaction
scenario, however, the more realistic estimates seem to be in 
agreement with the values found in the simulations. The situation
could be clarified when a reliable mass determination of DDO125 
is available.

\subsection{Origin of the extended HI--distribution around NGC4449}

   Another important question is related to the initial conditions
of the previous simulations, i.e.\ the origin of the extended
HI--distribution. In principle, there are several possibilities:

\noindent
{\it Idea 1. The HI comes from DDO125.}
In a previous encounter ($\epsilon<1$) the gas is stripped off 
from DDO125. Due to the long orbital period of e.g.\ 3.6 Gyr for 
$\epsilon=0.5$ the gas has enough time to settle down in a regular 
disc-like structure. However, the amount of HI gas in DDO125 is much
smaller than the gas seen around NGC4449. It seems to be unlikely that
DDO125 could have survived the loss of such a big fraction of its 
gas content without disruption. 

\noindent
{\it Idea 2. The HI comes from NGC4449.}
Though a star forming dwarf galaxy drives a galactic 
wind e.g.\ by means of type II supernovae, a detailed fine--tuning 
of the energy injection would be required to prevent 
a blow--out of the HI--gas, especially if it got enough energy
to form such an extended system. Moreover, the 
counter-rotation of the HI outside and inside the optical
radius of NGC4449 makes the scenario not very appealing.

\noindent
{\it Idea 3. The HI gas stems from a previous minor merger with
NGC4449}. Though a satellite galaxy can be
disrupted by the tidal field of NGC4449, it is very unlikely
that it is disrupted already far outside the galactic
center where the HI gas is detected.

\noindent
{\it Idea 4. The HI originates from the collapse of a dwarf companion.}
A dwarf galaxy forms in the vicinity of NGC4449.
By this it undergoes a strong collapse which leads to mass ejection 
due to violent relaxation. In principle, a mass loss of up to 30\% of
the companion could occur providing enough gas to explain the HI.
However, where is the dwarf today? Excluding an
unlikely radial orbit, the dwarf's orbit could in principle 
decay by dynamical friction. However, due to the large distance
to NGC4449 the decay time should be at least several orbital periods 
which exceeds the Hubble time.

\noindent
{\it Idea 5. Another major encounter.} Though NGC4449 is a member
of the loose Canes Venatici group, there seem to be
no close galaxy which has a radial velocity in agreement with
the rotational sense of the gaseous disc in NGC4449. So,
who is the partner?

%
\section{Summary and conclusions}
\label{conclusions}

 Several N-body methods are combined in order to develop
a method for the determination of the parameters of interacting
galaxies. This method is applied to the HI distribution
of NGC4449. 
In a first step the fast restricted N-body method is used
to confine a region in parameter space which reproduces the main
observed features. In a second step a genetic algorithm (also
using restricted N-body calculations) is employed
which allows for both, an automatic fit of observational
data even in a high-dimensional parameter space and/or 
a uniqueness test of a favoured parameter combination. For a
genetic algorithm one typically has to follow a population of (at least) 100
members for 100 generations in order to get a good fit provided the
data are sufficiently accurate. Missing single pixels do not
inhibit the parameter determination, as long as the key features are included
in the data. Since a typical restricted N-body simulation takes
a few CPU-seconds on a Sparc workstation, a whole fitting
procedure is finished after 3--6 CPU--hours.

  In the third step the results of the previous steps
are compared with detailed self--consistent N-body simulations.
In the case of NGC4449 they show that the restricted
N-body calculations are reliable models for this encounter. 
The comparison with the sticky particle models demonstrates
that the HI gas can be modeled without any restriction
by a purely stellar dynamical approach provided the encounter
is as weak as in the case of NGC4449 and DDO125.

 From the previous simulations I conclude that the extended
HI features observed in NGC4449 are created by an encounter
with DDO125. Prior to the encounter the HI gas formed
an extended, almost homogenous disc with a large scale length 
of about 30 kpc and a radial extension of 40 ($\pm10$) kpc.
The orientation of the disc is in agreement with the
orientation of the inner ellipsoidal HI distribution.
The orbital plane has an inclination
angle of about 40$^\circ$ ($\pm 10^\circ$). The eccentricity
is close to that of a parabolic encounter ($0.5<\epsilon<1.5$)
and the apocenter distance of the galaxies is about
25 ($\pm5$) kpc. The closest approach happened $\nat{3-4}{8}$ yr ago.
The mass ratio of both galaxies must
be approximately 0.2 (and cannot be smaller than 0.1).
 
{\bf Acknowledgements.} \hspace*{0.2cm}
The simulations were partly performed with the
GRAPE3af special purpose computer in Kiel (DFG Sp345/5).
The author is grateful to Deidre Hunter, Jay Gallagher, Uli Klein,
Sven Kohle and Hugo van Woerden for stimulating discussions about NGC4449.
I am also greatly indebted to Konrad Kuijken and John Dubinski, who made 
their program for the generation of initial particle distributions available
to the public, and Paul Charbonneau and Barry Knapp for providing their 
{\sc pikaia} code. Last, but not least, I want to thank Joachim
K\"oppen and Tim Freyer for a careful reading of the manuscript.

\section*{References}

\begin{description}
\itemsep-.8ex 
 
 \item Arp H., 1966, {\it Atlas of peculiar galaxies}, Pasadena, Caltech
          (also ApJS, 14,1)
 \item B\"ack Th., 1996, {\it Evolutionary Algorithms in Theory and Practice},
          Oxford Univ.\ Press, Oxford
 \item Bajaja E., Huchtmeier W.K., Klein U., 1994, A\&A, 285, 388
 \item Barnes J., 1988, ApJ, 331, 699
 \item Barnes J., Hernquist L., 1996, ApJ, 471, 115
 \item Barnes J., Hut P., 1986, Nature, 324, 446
 \item Bomans D., Chu Y.-H., 1997, AJ, 113, 1678
 \item Casoli F., Combes F., 1982, A\&A, 110, 287
 \item della Ceca R., Griffiths R.E., Heckman T.M., 1997, ApJ, 485, 581
 \item Charbonneau P., 1995, ApJS, 101, 309
 \item Crillon R., Monnet G., 1969, A\&A, 1, 449
 \item Ebneter K., Davis M., Jeske N., Stevens M., 1987, BAAS, 19, 681
 \item Fisher J.R., Tully R.B., 1981, ApJS, 47, 139
 \item Frenk C.S., White S.D.M., Davis M., Efstathiou G., 1988, ApJ, 327, 507 
 \item Gerhard O., 1981, MNRAS, 197, 179
 \item Goldberg D.E., 1989, {\it Genetic Algorithms in Search, Optimization,
         \& Machine Learning}, Addison-Wesley, Reading
 \item Hernquist L., 1987, ApJS, 64, 715
 \item Hernquist L., 1990, J. Comp.\ Phys., 87, 359
 \item Hernquist L., Katz N., 1989, ApJS, 70, 419
 \item Hernquist L., Ostriker J.P., 1992, ApJ, 386, 375
 \item Hill R.S., Home A.T., Smith A.M., Bruhweiler F.C., Cheng K.-P., 
          Hintzen P.M., Oliversen R.J., 1994, ApJ, 430, 568
 \item Holland J., 1975, {\it Adaptation in natural and artificial systems},
         Univ.\ of Michigan Press, Ann Arbor
 \item Holmberg E., 1941, ApJ, 94, 385
 \item Huang S., Carlberg R.G., 1997, ApJ, 480, 503
 \item Hunter D.A., Gallagher J.S., 1997, ApJ, 475, 65
 \item Hunter D.A., Gillet F.C., Gallagher J.S., Rice W.L., Low F.J., 1986, 
         ApJ, 303, 171
 \item Hunter D.A., Thronson H.A., 1996, ApJ, 461, 202
 \item Hunter D.A., Wilcots E.M., van Woerden H., Gallagher J.S.,
          Kohle S., 1998, ApJL, 495, L47
 \item Ibata R., Gilmore G., Irwin M., 1994, Nature, 370, 194
 \item Jungwiert B., 1997, private communication
 \item Kant I., 1755, in {\it Allgemeine Naturgeschichte und Theorie des
         Himmels oder Versuch von der Verfassung und dem mechanischen
         Ursprunge des ganzen Weltgeb\"audes nach Newtonischen
         Grunds\"atzen abgehandelt}, ed.\ H. Ebert, 1890, 
         Verlag Wilhelm Engelmann, Leipzig
 \item Karachentsev I.D., Drozdovsky I.O., 1998, A\&AS, 131, 1
 \item Klein U., Hummel E., Bomans D., Hopp U., 1996, A\&A, 313, 396
 \item Kohle S., Klein U., Henkel C., Hunter D.A., 1998, in
        Proc.\ of {\it The Magellanic Clouds and Other Dwarf Galaxies},
         ed.\ T. Richtler \&  J.M. Braun, Bad Honnef, see also
             http://www.astro.uni-bonn.de/~webgk/ws98/skohle\_p.html
 \item Kuijken K., Dubinski J., 1995, MNRAS, 277, 1341
 \item Noguchi M., 1988, A\&A, 203, 259
 \item Palou\v s J., Jungwiert B., Kopeck\'y J., 1993, A\&A, 274, 189
 \item Pfleiderer J., Siedentopf H., 1961, Zs.\ f.\ Ap., 51, 201
 \item Press W.H., Teukolsky S.A., Vetterling W.T., Flannery B.P., 1992,
        {\it Numerical Recipes in FORTRAN - The Art of Scientific Computing},
         Cambridge Univ.\ Press, Cambridge
 \item Quinn P.J., Hernquist L., Fullagar D.P., 1993, ApJ, 403, 74
 \item Rechenberg I., 1965, {\it Cybernetic solution path of an experimental
         problem}, Royal Aircraft Establishment, Library Translation No.\ 1122,
         Farnborough
 \item Rivolo A.V., Solomon P.M., 1987, in {\it Molecular Clouds in the 
     Milky Way and External Galaxies}, ed.\ R. L. Dickman et al., 
     Springer, p.\ 42
 \item Salo H., Laurikainen E., 1993, ApJ, 410, 586
 \item Sasaki M., Ohta K., Saito M., 1990, PASJ, 42, 361
 \item Schwefel H.-P., 1977, {\it Numerische Optimierung von 
     Computer--Modellen mittels der Evolutionsstrategie}, Birkh\"auser, Basel
 \item Sellwood J.A., 1980, A\&A, 89, 296
 \item Silk J., Wyse R.F.G., Shields G.A., 1987, ApJ, 322, L59
 \item Sugimoto D., Chikada Y., Makino J., Ito T., Ebisuzaki T.,
          Umemura M., 1990, Nature, 345, 33
 \item Theis Ch., Hensler G., 1993, A\&A, 280, 85
 \item Theis Ch., Kohle S., 1998, in
        Proc.\ of {\it The Magellanic Clouds and Other Dwarf Galaxies},
         ed.\ T. Richtler \&  J.M. Braun, Bad Honnef, see also
             http://www.astro.uni-bonn.de/~webgk/ws98/ctheis\_t.html
 \item Thomasson M., Donner K.J., 1993, A\&A, 272, 153
 \item T\'oth G., Ostriker J.P., 1992, ApJ, 389, 5
 \item Toomre A., 1977, in {\it The Evolution of Galaxies and Stellar
          Populations}, ed.\ B.M. Tinsley \& R.B. Larson, Yale Univ.\ Obs.\,
          New Haven
 \item Toomre A., Toomre J., 1972, ApJ, 178, 623 
 \item Tully R.B., Bottinelli L., Fisher J.R., Gouguenheim L., Sancisi R.,
          van Woerden H., 1978, A\&A, 63, 37
 \item van Woerden H., Bosma A., Mebold U., 1975, in {\it La Dynamique
          des Galaxies Spirales}, ed.\ L. Weliachew, p.\ 483
 \item Vel\'azquez H., White S.D.M., 1998, preprint, astro--ph/9809412
 \item Vorontsov--Velyaminov B.A., 1959, {\it Atlas and Catalog of
          Interacting Galaxies}, Moscow, Sternberg Institute
 \item Wahde M., 1998, A\&AS, 132, 417 
 \item Walker I.R., Mihos J.C., Hernquist L., 1996, ApJ, 460, 121
 \item White S.D.M., 1978, MNRAS, 184, 185
\end{description}

\end{document}